\documentclass[11pt,twoside]{article}
\usepackage{asp2010}
\usepackage{times}
\usepackage{natbib} 
\usepackage{amssymb}
\usepackage{textcomp}
\usepackage[normalem]{ulem}
\resetcounters

\begin{document}

\title{Reviewing the observational evidence against long-lived spiral arms in galaxies}
\author{Eric E. Mart\'inez-Garc\'ia $^1$ \& Ivanio Puerari $^1$
\affil{$^1$Instituto Nacional de Astrof\'isica, \'Optica y Electr\'onica (INAOE), Aptdo. Postal 51 y 216, 72000 Puebla, Pue., M\'exico}
}
\begin{abstract}

We review Foyle et al.~(2011) previous results, by applying
a Fourier intensity phases method to a nine object sample of galaxies.
It was found that two of the objects (NGC 628 and NGC 5194), with strong two-arm patterns,
present positive evidence for long-lived spirals.
Only one of the objects (NGC 3627) shows the contrary evidence.
As determined by an analysis of resolved mass maps,
the rest of the objects can not be included in the analysis because
they belong to flocculent and multi-arm type of spiral arms,
which are not described by density wave theory.

\end{abstract}

\section{Introduction}

As implied by the density wave (DW) theory, spiral patterns
should be thought of as dynamical features that are stationary in a corotating frame. 
If such spiral arms trigger star formation (SF), some
observational tracers for different stages of the SF
sequence should show a spatial ordering~\citep{foy11}.\footnote{
Besides the discussed SF sequence, azimuthal age/color gradients are also
expected across spiral arms. These gradients are indeed present in some objects as has been
probed by~\citet{gon96},~\citet{mar09a,mar09b}, and~\citet{mar11,mar13}}
From upstream to downstream in the corotating frame: dense H{\rm{I}} emission,
CO emission tracing molecular hydrogen gas, $24\micron$ emission tracing enshrouded SF, and UV emission tracing unobscured
young stars.~\citet{foy11} analyzed a sample of 12 nearby spiral galaxies, with grand-design, flocculent, and barred objects.
A cross-correlation method was adopted. Their findings argue for little evidence supporting
the DW scenario, even for the grand-design objects.

Our research involves the analysis of spiral galaxies in search for this
sequence of SF, by adopting the Fourier method of~\citet{pue97}.
As in~\citet{foy11}, the observational data consists on high-quality
maps of neutral gas (H{\rm{I}}) from THINGS~\citep{wal08}; 
molecular gas (CO) from HERACLES~\citep{ler09};
$24\micron$ emission from Spitzer~\citep[SINGS,][]{ken03};
FUV images from GALEX~\citep{gil07};
and $3.6\mu\mathrm{m}$ emission from Spitzer~\citep[SINGS,][]{ken03}.
Nine nearby spiral galaxies have common data in all the aforementioned surveys and were selected
(NGC 628, 2403, 2841, 3351, 3521, 3627, 5055, 5194 and NGC 6946).
With the exception of NGC 6946, all the objects also have also cross-sections in SDSS~\citep{yor00}.

\section{Candidates that may host density waves}

Not all types of spirals can be explained by DW theory.
Only the symmetric grand-design objects may harbor
mass structure with nearly constant patterns speeds, i.e., density waves.
Thereby multi-armed and flocculent spirals are naturally excluded
as stellar DW candidates~\citep{efr11,mar13}. In order to discern whether these structures are present in the disks
of our sample of galaxies, we adopt the following method.
Resolved maps of stellar mass were obtained 
by extending the~\citet{zib09} method to the $3.6\micron$-band of IRAC-Spitzer.
This was done because these near-infrared (NIR) images are available for our entire sample of objects,
in addition to its quality. The dust emission model of~\citet{dac08} was adopted
to include dust radiation into account. A Monte Carlo library of stellar population synthesis (SPS) models,
and dust radiation models was obtained as in~\citet{zib09}. An energy balance
condition~\citep{dac08} was applied so that all the radiation absorbed in the ISM, and stellar birth clouds,
is re-radiated in the infrared for wavelengths longer than $2.5\micron$.
A look-up table was then derived for the $(g-i)$ color, the $(i-3.6\micron)$ color, and the mean $M/L_{(3.6\micron)}$ ratio.
This look-up table was compared with SDSS\footnote{The Johnson's $V$ and $I$ filters were used for NGC 6946.}
and $3.6\micron$ photometry in a pixel-by-pixel basis,
thereby obtaining a map of the $M/L_{(3.6\micron)}$ ratio. These maps were transformed to absolute
mass maps with distances obtained from the literature.

\begin{figure}[t]


\centering
\includegraphics[scale=0.4]{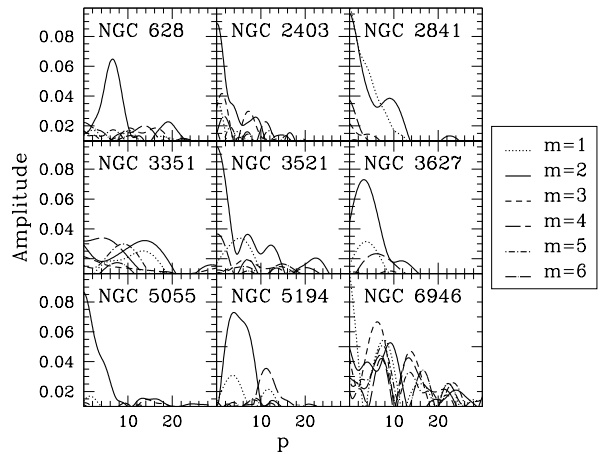}

  \caption{Amplitudes vs. $p$ frequencies (absolute values), obtained with 2D FFT methods,
   for the resolved mass maps in our sample of objects.
   The $p$ frequency is the one associated with the pitch angle of the object~\citep[see, e.g.,][]{sar94}.
    \label{fig1}}
\end{figure}

The two-dimensional mass maps were then analyzed to select those with a strong two-armed
spiral pattern ($m=2$). This was done with 2D FFT methods, the same used for pitch angle
measurements~\citep[e.g.,][]{con82,pue92,sar94,dav12}. The results are shown in figure~\ref{fig1},
where we can see that only three objects (NGC 628, NGC 3627 and NGC 5194) show signs of strong $m=2$ spiral patterns.
These objects were considered for the analysis of azimuthal phases of intensity as described below.

\section{Fourier method of azimuthal phases}

~\citet{pue97} method is based on computing the Fourier transform of the form

\begin{equation}
    \hat{f}(m)=\int_{-\pi}^{\pi} I_{R}(\theta) e^{-im \theta}\, d\theta,
\end{equation}

\noindent where $I_{R}$ is the intensity of radiation, and the phase

\begin{equation}
    \Phi = \tan^{-1} \left\{ \frac{\mathrm{Re}[\hat{f}(m)]}{\mathrm{Im}[\hat{f}(m)]} \right\},    
\end{equation}

\noindent for two-armed spiral modes, i.e., $m=2$.
This implies that twice the period of the analyzed signal is contained within a $2\pi$ radians interval,
i.e., a $\pi$ radians symmetry. The azimuthal phases were computed for the H{\rm{I}}, CO, $24\micron$, and FUV data.
All data were degraded to the CO~\citep[Heracles,][]{ler09} resolution.
For shock-induced star formation in a DW scenario~\citep{rob69}, before the corotation,
we expect to find a higher phase value for the H{\rm{I}} data,
and lower phase values for the CO, $24\micron$, and FUV data, in an ordered sequence.
The phases most coincide near corotation, and invert their order afterwards. 

\section{Results}

\begin{figure}[t]

\centering
\includegraphics[scale=0.75]{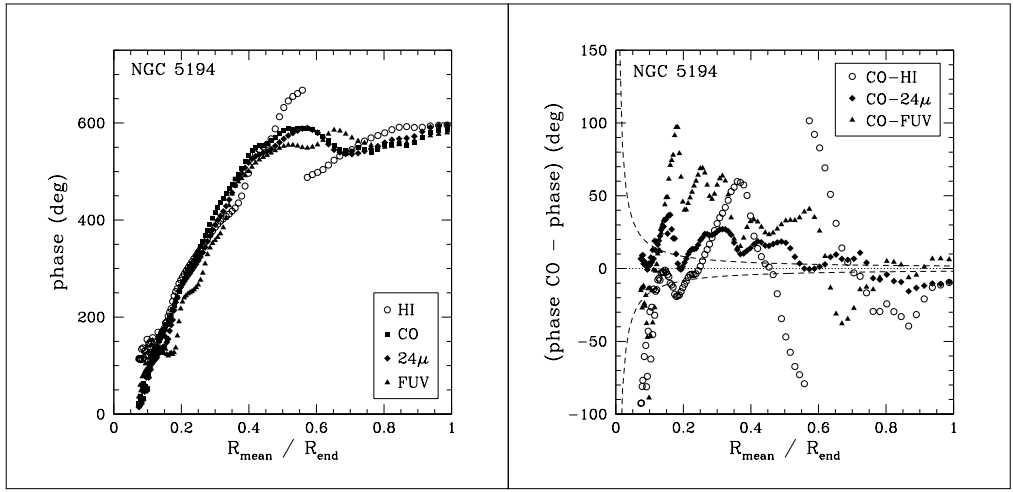}

  \caption{Azimuthal intensity phases for NGC 5194. Left: Azimuthal phases ($y$-axis), vs.,
   normalized radius $R_{\rm{mean}}/R_{\rm{end}}$ ($x$-axis), where
   $R_{\rm{end}}$ indicates the end of the spirals in the $3.6\micron$ image.
   Right: Differences between the phases in CO and other tracers ($y$-axis),
   vs., normalized radius $R_{\rm{mean}}/R_{\rm{end}}$ ($x$-axis). The long-dashed line indicates the resolution
   of the data.
   \label{fig2}}
\end{figure}

For the three objects analyzed, it was found that none of the H{\rm{I}}, and FUV phase plots,
show the expected sequence if DW's are indeed present in the disk. Although, if the H{\rm{I}}
emission is contaminated by the gas photodissociated by recently formed stars,
it would not trace the compressed gas~\citep{lou13}. 
For NGC 628 and NGC 5194, it was found that the CO and $24\micron$
emission show the expected sequence for the DW scenario.
For NGC 628, the CO-$24\micron$ phase shift
coincides with the expectations up to the place where the spiral arms seem to end,
as inferred from the mass map. For NGC 5194, the phase shifts agree with a DW scenario for
most of its spiral extension~(see figure~\ref{fig2}).
No consistent phases were found for NGC 3627.

\section{Conclusions}

We analyzed a sample of nine objects, with available data in
H{\rm{I}}, CO, $24\micron$, and FUV. Based on SDSS, and $3.6\micron$ photometry,
resolved mass maps were obtained.
In these mass maps, three objects (NGC 628, NGC 3627, and
NGC 5195) show a strong spiral pattern with $m=2$.
After applying a Fourier method to calculate azimuthal
phases~\citep{pue97}, two of the objects, NGC 628, and NGC 5194,
showed indications of a corotation (at least for the
phase shift between CO and $24\micron$).
This is clear sign of a dominant pattern speed with a constant
value, i.e., long-lived spirals.

\acknowledgements EMG acknowledges support from the mexican institution CONACYT.

\bibliography{MartinezgarciaE}

\end{document}